\documentclass{ws-mplb}
\usepackage{amsfonts,amssymb,amsmath}
\usepackage{graphicx}
\usepackage[super,sort,compress]{cite} % sort => 1,2,3,5; nosort => 1,3,2,5; compress => 1-3,5; nocompress => 1,2,3,5

\begin{document}

\markboth{R. Mia and A.K. Paul}{New exact solutions to the GSWWE}

%%%%%%%%%%%%%%%%%%%%% Publisher's Area please ignore %%%%%%%%%%%%%%%
%
\catchline{}{}{}{}{}
%
%%%%%%%%%%%%%%%%%%%%%%%%%%%%%%%%%%%%%%%%%%%%%%%%%%%%%%%%%%%%%%%%%%%%

\title{{New exact solutions to the generalized shallow water wave equation}
}

\author{\footnotesize Rajib Mia\footnote{Corresponding author.}}

\address{School of Applied Sciences, Department of Mathematics, Kalinga Institute of Industrial Technology
Deemed to be University,\\
 Bhubaneswar-751024, Odisha, India
\footnote{State completely without abbreviations, the
affiliation and mailing address, including country.}\\
rajibmia.90@gmail.com}

\author{Arjun Kumar Paul}

\address{School of Applied Sciences, Department of Mathematics, Kalinga Institute of Industrial Technology
Deemed to be University,\\
 Bhubaneswar-751024, Odisha, India\\
apaulfma@kiit.ac.in}

\maketitle

\begin{history}
\received{(Day Month Year)}
\revised{(Day Month Year)}
\end{history}

\begin{abstract}
In this work, we study the generalized shallow water wave equation to obtain novel solitary wave solutions. The application of this non-linear model can be found in tidal waves, weather simulations, tsunami prediction, river and irrigation flows, etc. To obtain the new exact solutions of the considered model, we have applied a novel analytical technique namely $\left(\frac{G'}{G'+G+A}\right)$--expansion method. Using the aforementioned method and computational software, we have obtained different kinds of periodic and singular solitary wave solutions of the generalized shallow water wave equation. The obtained solutions are exponential function and trigonometric function solutions. Using 2-D and 3-D plots of the wave solutions, the dynamic behaviors of the developed solutions are displayed. The retrieved solutions validated the effectiveness and robustness of the proposed technique.
\end{abstract}

\keywords{Non-linear evolution equations; GSWWE; Exact solutions; Soliton solutions.}
\section{Introduction}
%Regarding nonlinear equation:\\
Non-linear partial differential equations (NPDEs) can be extensively used to model a variety of physical phenomena including electromagnetic waves, solid mechanics, optical fibres, biomedical sciences, plasma physics,  signal processing, fluid dynamics, and quantum mechanics, etc.
%In this study, we deal with 
In this study, we will deal with the following generalized shallow water wave equation (GSWWE) \cite{selima2016nonlinear,miah2020abundant}
\begin{eqnarray}
v_{xxxt}+\alpha v_x v_{xt}+\beta v_t v_{xx}-\gamma v_{xx}-v_{xt}=0, \label{eq:1}
\end{eqnarray}
where $\alpha$, $\beta$ and $\gamma$ are non-zero constants.
%Regarding Generalized SWWE:\\
The applications of the shallow water wave equations (SWWEs) can be found in tidal waves, weather simulations, tsunami prediction, river and irrigation flows etc. For this wide range of applications, many researchers studied the shallow water wave equations using different analytical techniques  \cite{elwakil2003exact,kumar2019some,liu2019breather,chu2021diverse,dawod2023breather}. 

%Regarding several methods:\\
There are several researchers who have developed and implemented various analytical techniques to extract the exact solutions of the non-linear models involving NPDEs. For instance  Lie symmetry analysis \cite{liu2009lie,sharmainvariance,ray2017conservation,ali2022lie},  the generalized Kudryashov approach \cite{barman2021harmonizing,akbar2022dynamical,abdou2023plenteous},  Paul--Painlev$\acute{e}$ approach \cite{singh2023new}, 
$\left(\frac{G'}{G},\frac{1}{G}\right)$-expansion method \cite{miah2023study}, the Riccati equation method \cite{mohammed2022analytical,ali2023investigation}, the
F-expansion principle \cite{yildirim2021optical,ali2023investigation}, the new extended auxiliary equation method \cite{rizvi2020optical,mathanaranjan2023chirped}, the Jacobi elliptic
expansion method \cite{liu2001jacobi,ahmed2021optical,tarla2022dynamic}, the tanh-function method \cite{evans2005tanh},  $\left(\frac{G'}{G}\right)$-expansion method \cite{bekir2008application,alam2013novel,mohammed2021exact}, the exponential function technique \cite{javeed2020soliton,zulfiqar2020soliton,sajid2022implementation}, the Sardar sub--equation method \cite{asjad2022traveling,rehman2023analysis},   
the sine-Gordon expansion method \cite{kumar2017sine,baskonus2019new,ali2020new,abdou2023plenteous}, the improved sub-equation method \cite{ozkan2022analytical,akar2023exact},  the functional variable method \cite{ccevikel2012procedure} and so on.

%Regarding Several Solutions
%Different types of analytical solutions of  partial differential equations  have been reported in various literature. For instance, multi-soliton, multi-M-lump, and hybrid soliton are reported by Ismael \citep{ismael20233+}.

%Regarding recent work of nonlinear model:\\
Different types of analytical solutions of  partial differential equations  have been reported in various literature. For instance, multi-soliton, multi-M-lump, and hybrid soliton solutions are obtained by Ismael \cite{ismael20233+}.
In recent studies, the soliton solutions of shallow water waves and superconductivity models using the rational $\left(\frac{G'}{G}\right)$-expansion approach are obtained by Akbar et al.\cite{akbar2023new}. Two significant methods, the improved Bernoulli sub-equation function and the new auxiliary equation
methods are used to extract exact solutions to the Kadomtsev--Petviashvili equation \cite{islam2023solitary}. Utilizing the generalized Riccati-equation mapping technique, the generalized complex coupled Schr$\ddot{\mbox{o}}$dinger--Boussinesq are explored, in addition to several newly developed solitons and dynamic waveform structures by Kumar and Mann \cite{kumar2023variety}.

%Regarding other methods and solutions to the proposed model

%The purpose of the current work
The purpose of this study is to utilize a recently developed technique namely, $\left(\frac{G'}{G'+G+A}\right)$ -expansion method to extract novel analytical solutions of the generalized shallow water wave equation. The $\left(\frac{G'}{G'+G+A}\right)$ -expansion approach has only been used in a few papers \cite{hong2019g,tripathy2020exact,khaliq2022some,mia2023new,khaliq2023new}.
We have obtained some new analytical solutions for the GSWWE. The Mathematica program is used to display the 2-D and 3-D graphical representations of the discovered solutions. The wave solutions including kink, singular periodic, and soliton solutions are reported in this current study. According to the knowledge we have, the derived solutions are original and have not yet been revealed. 
In addition, the simulations and observations of many characteristics of the waves would be highly valuable in gathering more knowledge about the GSWWE.

%Organization of the paper
The organization of this paper is as follows: We have outlined the description and key phases of the suggested approach in Section \ref{sec:2}. In Section \ref{sec:3}, we have described the use of the suggested method to derive the analytical solutions to the generalized shallow water wave equation.
 In  Section \ref{sec:4}, simulation results for the specific parameter values are discussed. Finally, Section \ref{sec:5} contains conclusions.
\section{Description of the proposed technique} \label{sec:2}
In this section, the steps of $\left( \frac{G'}{G'+G+A}\right) $ expansion method are discussed.
Let us assume the general non-linear PDE of the following type
\begin{eqnarray}
F_1(v, v_x, v_y, v_t, v_{xx}, v_{yy}, v_{xy}, v_{xt}, \dots)=0, \label{eq:2}
\end{eqnarray}
where $v$ is a dependent variable and $x, y$ and $t$ are independent variables, $F_1$ is a polynomial in  $v$ and its  derivatives which involved non-linear and higher order derivative terms.
The following are the key phases of the suggested
 technique:\\
\textbf{Step 1}:  We assume that $\zeta=n x+m y-\omega t$, where $\omega$, $m$ and $n$ are constants. Let us consider the solution of Eq. (\ref{eq:2}) in new travelling  variable as 
\begin{eqnarray}
v(x, y, t)=u(\zeta). \label{eq:3}
\end{eqnarray}
Utilizing the above transformations in Eq. (\ref{eq:2}), the following ODE is obtained:
\begin{eqnarray}
F_2[u(\zeta), u'(\zeta), u''(\zeta), u'''(\zeta), \dots]=0. \label{eq:4}
\end{eqnarray}
%where $u'=\frac{du}{d\zeta},~u''=\frac{d^2u}{d\zeta^2},\dots$.\\

\textbf{Step 2}:
We assume the solution of Eq. (\ref{eq:4}) in the following form
\begin{eqnarray}
u(\zeta)=\sum_{k=0}^{N} \alpha_k \left(\frac{G'}{G'+G+A}\right)^{k}, \label{eq:5}
\end{eqnarray} 
where $G=G(\zeta)$ is the solution of the following second order linear ODE
\begin{eqnarray}
G''+B G'+C G+A C=0. \label{eq:6}
\end{eqnarray}
where $A, B, C$ and $\alpha_k (k=0, 1, 2, \dots, N)$ are constants.
Additionally, the polynomial's degree, $N$, can be calculated by using the homogeneous balancing method, which involves finding the homogeneous balance between the derivatives of the highest order and the highest order non-linear terms in Eq. (\ref{eq:4}).
Furthermore, a set of algebraic equations that will result from the suggested technique can be used to calculate the coefficients
 $\alpha_k~(k=0, 1, 2, \dots, N )$.  Then the function $G(\zeta)$ can be evaluated by putting the values of $A, B$ and $C$ in the SLODE given in Eq. (\ref{eq:6}).\\
\textbf{Step-3:}
Then, we can obtain the polynomial of  $\left( \frac{G'}{G'+G+A}\right)$ by inserting Eq. (\ref{eq:5}) into Eq. (\ref{eq:4}). After that, one can arrive to a system of algebraic equations for $\eta$, $\alpha_j$, $A, B,$ and $C$ by comparing the coefficients of equivalent powers of  $\left( \frac{G'}{G'+G+A}\right)$.

\section{Application of the $\left( \frac{G'}{G'+G+A}\right)$--expansion scheme to the GSWWE } \label{sec:3}
This current section is devoted to the application of the proposed technique to the GSWWE given in Eq. (\ref{eq:1}) for extracting the new exact closed-form solutions. For our considered model, using the traveling wave transformation $v(x, t)=u(\zeta),~ \zeta=x-\omega t$, the GSWWE in Eq. (\ref{eq:1}) converts into the following equation
\begin{eqnarray}
-\omega u^{iv}-(\alpha+\beta)\omega u' u''+(\omega-\gamma)u''=0. \label{eq:7}
\end{eqnarray}
Integrating the Eq. (\ref{eq:7}), we obtain
\begin{eqnarray}
 u'''+\frac{1}{2}(\alpha+\beta) u'^2+\eta u'=0, \label{eq:8}
\end{eqnarray}
where $\eta=\frac{\gamma}{\omega} -1.$
Let us suppose that $u'(\zeta)=\Phi (\zeta)$ and then the above equation (\ref{eq:8}) reduces to the following second order ODE
\begin{eqnarray}
\Phi''+\frac{1}{2}(\alpha+\beta) \Phi^2+\eta \Phi=0.\label{eq:9}
\end{eqnarray}

On balancing the largest order derivative term and the largest order no-linear term of Eq. (\ref{eq:9}), we obtain $N=2$. Then the solution of Eq. (\ref{eq:9}) is given by
\begin{eqnarray}
\Phi(\zeta)=\sum_{k=0}^{2} \alpha_k \left(\frac{G'}{G'+G+A}\right)^{k}.  \label{eq:10} 
\end{eqnarray} 
Inserting Eq. (\ref{eq:10}) into Eq. (\ref{eq:9}) and collecting the coefficients of similar exponents of $\left(\frac{G'}{G'+G+A}\right)$ and equating them, we get 
\begin{eqnarray}
\frac{\alpha _0^2 \beta }{2}+\alpha _0 \eta +\frac{\alpha  \alpha _0^2}{2}+\alpha _1 B C-2 \alpha _1 C^2+2 \alpha _2 C^2=0,\label{eq:11}
\end{eqnarray}
\begin{eqnarray}
\alpha _0 \alpha _1 \beta +\alpha _1 \eta +\alpha  \alpha _0 \alpha _1+\alpha _1 B^2-6 \alpha _1 B C+6 \alpha _2 B C+6 \alpha _1 C^2\nonumber\\
-12 \alpha _2 C^2+2 \alpha _1 C=0,\label{eq:12}
\end{eqnarray}
\begin{eqnarray}
\frac{\alpha _1^2 \beta }{2}+\alpha _0 \alpha _2 \beta +\alpha _2 \eta +\frac{\alpha  \alpha _1^2}{2}+\alpha  \alpha _0 \alpha _2-3 \alpha _1 B^2+4 \alpha _2 B^2+3 \alpha _1 B\nonumber\\\label{eq:13}
+9 \alpha _1 B C-24 \alpha _2 B C-6 \alpha _1 C^2+24 \alpha _2 C^2-6 \alpha _1 C+8 \alpha _2 C=0,
\end{eqnarray}
\begin{eqnarray}
\alpha _1 \alpha _2 \beta +2 \alpha _1+\alpha  \alpha _1 \alpha _2+2 \alpha _1 B^2-10 \alpha _2 B^2-4 \alpha _1 B+10 \alpha _2 B\nonumber\\\label{eq:14}
-4 \alpha _1 B C+30 \alpha _2 B C+2 \alpha _1 C^2-20 \alpha _2 C^2+4 \alpha _1 C-20 \alpha _2 C=0,
\end{eqnarray}
\begin{eqnarray}
\frac{\alpha _2^2 \beta }{2}+\frac{\alpha  \alpha _2^2}{2}+6 \alpha _2+6 \alpha _2 B^2-12 \alpha _2 B-12 \alpha _2 B C+6 \alpha _2 C^2\nonumber\\
+12 \alpha _2 C=0.\label{eq:15}
\end{eqnarray}
 Using the Mathematica computational software program, solving the above system of equations (Eqs. (\ref{eq:11})--(\ref{eq:15})), the following two set solutions are obtained:

\textbf{SET--1}:
\begin{eqnarray}
\eta = 4 C-B^2, ~ \alpha _0= -\frac{12 \left(-B C+C^2+C\right)}{\alpha +\beta },\nonumber\\ 
\alpha _1= \frac{12 \left(B^2-3 B C-B+2 C^2+2 C\right)}{\alpha +\beta }, ~\alpha _2= -\frac{12 (B-C-1)^2}{\alpha +\beta }.
\end{eqnarray}

\textbf{SET--2}:
\begin{eqnarray}
\eta = B^2-4 C, ~ \alpha _0= -\frac{2 \left(B^2-6 B C+6 C^2+2 C\right)}{\alpha +\beta },\nonumber\\
 \alpha _1= \frac{12 \left(B^2-3 B C-B+2 C^2+2 C\right)}{\alpha +\beta }, ~\alpha _2= -\frac{12 (B-C-1)^2}{\alpha +\beta }.
\end{eqnarray}

 \vspace*{0.8cm}
 In the case of \textbf{SET--1}, we obtain the following analytical wave solutions:\\
 Case-I: When $\Omega=B^2-4C>0$
 \begin{eqnarray}\label{eq:16}
&&\Phi_{11}(x, t)= -\frac{12 \left(-B C+C^2+C\right)}{\alpha +\beta }\nonumber\\
&&~~~ +\frac{12 \left(B^2-3 B C-B+2 C^2+2 C\right)}{\alpha +\beta }\times\\
&&~~~\left[ \frac{k_2 \left(B-\sqrt{\Omega }\right) e^{\zeta  \sqrt{\Omega }}+k_1 \left(B+\sqrt{\Omega }\right)}{k_2 \left(B-\sqrt{\Omega }-2\right) e^{\zeta  \sqrt{\Omega }}+k_1 \left(B+\sqrt{\Omega }-2\right)}\right] \nonumber\\
&& ~~~-\frac{12 (B-C-1)^2}{\alpha +\beta }\times\nonumber\\ 
&&\left[ \frac{k_2 \left(B-\sqrt{\Omega }\right) e^{\zeta  \sqrt{\Omega }}+k_1 \left(B+\sqrt{\Omega }\right)}{k_2 \left(B-\sqrt{\Omega }-2\right) e^{\zeta  \sqrt{\Omega }}+k_1 \left(B+\sqrt{\Omega }-2\right)}\right] ^2.\nonumber
 \end{eqnarray}
 
  \vspace*{0.8cm}
 Case-II: When $\Omega=B^2-4C<0$
 \begin{eqnarray}
 \tiny
&&\Phi_{12}(x, t)=  -\frac{12 \left(-B C+C^2+C\right)}{\alpha +\beta }\nonumber\label{eq:17}\\
&&  +\frac{12 \left(B^2-3 B C-B+2 C^2+2 C\right)}{\alpha +\beta })\times\\
&&\left[ \frac{k_1 \left(\sqrt{-\Omega } \sin \psi+B \cos \psi\right)+k_2 \left(B \sin \psi-\sqrt{-\Omega } \cos \psi\right)}{k_1 \left((B-2) \cos \psi+\sqrt{-\Omega } \sin \psi\right)+k_2 \left((B-2) \sin \psi-\sqrt{-\Omega } \cos \psi\right)}\right] \nonumber\\
&& -\frac{12 (B-C-1)^2}{\alpha +\beta }\times\nonumber\\
&&\left[ \frac{k_1 \left(B \cos \psi+\sqrt{-\Omega } \sin \psi\right)+k_2 \left(B \sin \psi-\sqrt{-\Omega } \cos \psi\right)}{k_1 \left((B-2) \cos \psi+\sqrt{-\Omega } \sin \psi\right)+k_2 \left((B-2) \sin \psi-\sqrt{-\Omega } \cos \psi\right)}\right]^2,\nonumber
 \end{eqnarray}
 where $\psi=\frac{\zeta \sqrt{-\Omega}}{2}$.
 
 \vspace*{0.8cm}
 For \textbf{SET--2} the exact traveling wave solutions:\\
 Case-I: When $\Omega=B^2-4C>0$
  \begin{eqnarray}
&&\Phi_{21}(x, t)=   -\frac{2 \left(B^2-6 B C+6 C^2+2 C\right)}{\alpha +\beta }\nonumber\label{eq:18}\\
&&+\frac{12 \left(B^2-3 B C-B+2 C^2+2 C\right)}{\alpha +\beta }\times\\
&&\left[ \frac{k_2 \left(B-\sqrt{\Omega }\right) e^{\zeta  \sqrt{\Omega }}+k_1 \left(B+\sqrt{\Omega }\right)}{k_2 \left(B-\sqrt{\Omega }-2\right) e^{\zeta  \sqrt{\Omega }}+k_1 \left(B+\sqrt{\Omega }-2\right)}\right] \nonumber\\
&& -\frac{12 (B-C-1)^2}{\alpha +\beta }\times\nonumber\\
&&\left[ \frac{k_2 \left(B-\sqrt{\Omega }\right) e^{\zeta  \sqrt{\Omega }}+k_1 \left(B+\sqrt{\Omega }\right)}{k_2 \left(B-\sqrt{\Omega }-2\right) e^{\zeta  \sqrt{\Omega }}+k_1 \left(B+\sqrt{\Omega }-2\right)}\right] ^2.\nonumber
 \end{eqnarray}
 Case-II: When $\Omega=B^2-4C<0$
  \begin{eqnarray}
&&\Phi_{22}(x, t)=  -\frac{2 \left(B^2-6 B C+6 C^2+2 C\right)}{\alpha +\beta }\nonumber\\
&& +\frac{12 \left(B^2-3 B C-B+2 C^2+2 C\right)}{\alpha +\beta }\times\label{eq:19}\\
&&\left[\frac{k_1 \left(B \cos \psi+\sqrt{-\Omega } \sin \psi\right)+k_2 \left(B \sin \psi-\sqrt{-\Omega } \cos \psi\right)}{k_1 \left((B-2) \cos \psi+\sqrt{-\Omega } \sin \psi\right)+k_2 \left((B-2) \sin \psi-\sqrt{-\Omega } \cos \psi\right)}\right]\nonumber \\
&&  -\frac{12 (B-C-1)^2}{\alpha +\beta }\times\nonumber\\
&&\left[ \frac{k_1 \left(B \cos \psi+\sqrt{-\Omega } \sin \psi\right)+k_2 \left(B \sin \psi-\sqrt{-\Omega } \cos \psi\right)}{k_1 \left((B-2) \cos \psi+\sqrt{-\Omega } \sin \psi\right)+k_2 \left((B-2) \sin \psi-\sqrt{-\Omega } \cos \psi\right)}\right]^2.\nonumber
 \end{eqnarray}
\section{Simulation results and discussion} \label{sec:4}
In this part, we go over the graphical representations of the derived traveling wave solutions of the GSWWE. On integration of solutions $\Phi_{11}, \Phi_{12}, \Phi_{21}$ and $\Phi_{22}$ given in Eqs. (\ref{eq:16})--(\ref{eq:19}) with respect to $\zeta$, one can achieved the required solutions. We have shown four distinct solutions obtained by the proposed method in 3D and 2D plots. For all the figures, we have taken $\alpha=\beta=\gamma=1$.
The first soliton solution for SET-1 with $\Omega>0$ is depicted in Fig. \ref{fig:1} for parameters values $ B = 1, C = 0.1,  k_1 =1, k_2 = 1$. In Fig. \ref{fig:1},  (I) shows the 3D plot and  (II) shows the 2D plot of the solution corresponding to $\Phi_{11}(x, t)$. 
In Fig. {\ref{fig:2}, we have presented the second solution corresponding to $\Phi_{12}$ for SET-2 with $\Omega<0$ for parameters values $  B = 1, C = 1.1,  k_1 =1, k_2 = 1$. From the 3D and 2D plots of this solution, it is clear that the wave solution is singular and periodic.
The third solution (soliton) corresponding to $\Phi_{21}$ for SET-1 with $\Omega>0$ is shown in Fig. \ref{fig:3} for parameters values $ B =1, C = 0.1,  k_1 =1, k_2 = 1$. For SET-2 and $\Omega<0$, the fourth wave solution corresponding to $\Phi_{22}$ is shown in Fig. \ref{fig:4} for parameters values $ B =1, C = 1.1,  k_1 =1.5, k_2 = 1$. In this case, it is observed that the obtained solution is singular and periodic.

%%%%%%%%%%%%%%%%%%Figures%%%%%%%%%%%%%%%%%%%%%%%%%%5
                     %Fig 1: Set 1,  $B^2-4C>0$, 
\begin{figure*}
\includegraphics[height=0.40\textwidth]{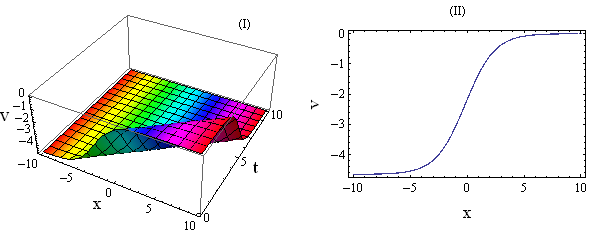}
\caption{3D and 2D graphical representation of solution corresponding to Eq. (\ref{eq:16}) for  $ B = 1, C = 0.1, k_1=k_2 = 1$.\label{fig:1}}
\end{figure*}

                    % Fig 2: Set 1, $B^2 - 4 C < 0$, 
\begin{figure*}
\includegraphics[height=0.40\textwidth]{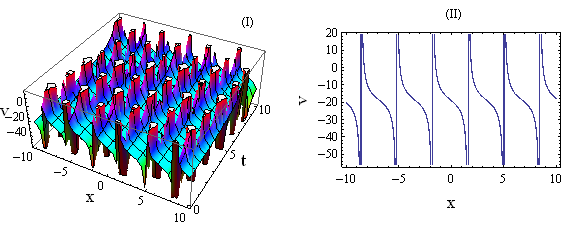}
\caption{3D and 2D graphical representation of solution corresponding to Eq. (\ref{eq:17}) for  $ B = 1, C = 1.1, k_1=k_2 = 1$. \label{fig:2}}
\end{figure*}

                %Fig 3: Set 2, $B^2 - 4 C >0$,
\begin{figure*}
\includegraphics[height=0.4\textwidth]{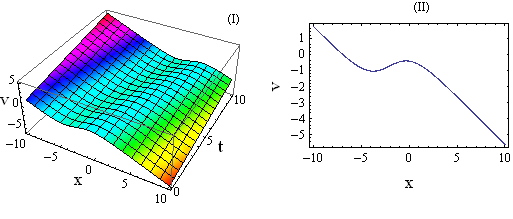}
\caption{3D and 2D graphical representation of solution corresponding to Eq. (\ref{eq:18}) for $B =1, C = 0.1, k_1 =k_2 = 1$. \label{fig:3}}
\end{figure*}

                  %Fig 4: Set 2, $B^2 - 4 C <0$,  
\begin{figure*}
\includegraphics[height=0.4\textwidth]{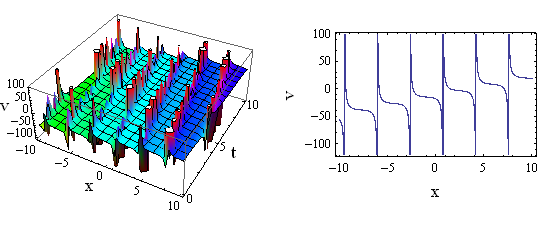}
\caption{3D and 2D graphical representation of solution corresponding to Eq. (\ref{eq:19}) for $ B =1, C = 1.1, k_1 =1.5, k_2 = 1$. \label{fig:4}}
\end{figure*}

\section{Conclusions} \label{sec:5}
In this work, we have implemented one most recently established novel analytical method to extract the exact closed-form solutions to the generalized shallow water wave equation (GSWWE). 
We have given the simulations of the derived solutions with the help of the Mathematica program. The 3D and 2D plots of four obtained solutions corresponding to $\Phi_{11}, \Phi_{12},\Phi_{21}$ and $\Phi_{22}$  are shown in Figs. \ref{fig:1}-\ref{fig:4} respectively for specific values of the parameters.
The findings of our study show that the extracted solutions are solitary wave solutions. Types of these solitary waves include kink solutions, singular periodic solutions, and the soliton wave solution.
Furthermore, we anticipate that the method for direct expansions described here should also be capable of examining a broad variety of non-linear models involving PDEs in mathematical physics and engineering.

%\begin{figure}[th]
%\centerline{\includegraphics[width=5cm]{mplbf1}}
%\vspace*{8pt}
%\caption{A schematic illustration of dissociative
%recombination. The direct mechanism,
%4m$^2_\pi$ is initiated when the molecular ion S$_{\rm L}$ captures an
%electron with kinetic energy.\label{f1}}
%\end{figure}

\hfill\eject

%\section{References}
%\section*{Acknowledgments}
     %This section should come before the References. Funding
      %information may also be included here.

%\appendix

%\section{Heading of Appendix}

%\section*{References}
\bibliographystyle{ws-mplb} 
\bibliography{mybibfile}

\end{document}